\documentclass[aps,prl,reprint,superscriptaddress,showpacs]{revtex4-1}
\usepackage{amsmath}
\usepackage{graphicx}
\usepackage{natbib}

\begin{document}

\title{Tuning of Near- and Far-Field Properties of All-dielectric Dimer Nanoantennas via Ultrafast Electron-Hole Plasma Photoexcitation}

\author{Denis G. Baranov}
\email[]{denis.baranov@phystech.edu}
\affiliation{Moscow Institute of Physics and Technology, 9 Institutskiy per., Dolgoprudny 141700, Russia}

\author{Sergey V. Makarov}
\affiliation{ITMO University, Saint Petersburg, Russia}

\author{Alexander E. Krasnok}
\affiliation{ITMO University, Saint Petersburg, Russia}

\author{Pavel A. Belov}
\affiliation{ITMO University, Saint Petersburg, Russia}

\author{Andrea Alu}
\affiliation{Department of Electrical and Computer Engineering, The University of Texas at Austin, Austin, Texas 78712, USA}

\date{\today}

\begin{abstract}
Achievement of all-optical ultrafast signal modulation and routing by a low-loss nanodevice is a crucial step towards an ultracompact optical chip with high performance. Here, we propose a specifically designed silicon dimer nanoantenna, which is tunable via  photoexcitation of dense electron-hole plasma with ultrafast relaxation rate. Basing on this concept, we demonstrate the effect of beam steering up to 20 degrees via simple variation of incident intensity, being suitable for ultrafast light routing in an optical chip. The effect is demonstrated both in the visible and near-IR spectral regions for silicon and germanium based nanoantennas. We also reveal the effect of electron-hole plasma photoexcitation on local density of states (LDOS) in the dimer gap and find that the orientation averaged LDOS can be altered by 50\%, whereas modification of the projected LDOS can be even more dramatic – almost 500\% for transverse dipole orientation. Moreover, our analytical model sheds light on transient dynamics of the studied nonlinear nanoantennas, yielding all temporal characteristics of the proposed ultrafast nanodevice. The proposed concept paves the ways to creation of low-loss, ultrafast, and compact devices for optical signal modulation and routing.
\end{abstract}

\maketitle

\section{Introduction}
The ability to control scattering of light by nanostructures is crucial for the development of functional nanodevices for data processing and information transfer. Nanoantenna switching usually implies control of extinction and absorption cross-sections~\cite{Engheta2008,AluCloak}, scattering patterns~\cite{Kall,Shegai2,Knoester,Trimer,Dorpe} and near field distribution~\cite{Volpe,Spin}.
Such alteration of optical properties can be achieved via an external impact, e.g. electro-optic effect~\cite{Load1,goldAPL}, magneto-optic effect~\cite{MO}, thermo-optical effect~\cite{Notomi} or carriers injection~\cite{Muskens}.

Alternatively, one may utilize the nonlinear response of the structure materials and control scattering with intensity of incident light~\cite{Alu,ITO,Lippitz16}. 
Noble metals offer Kerr nonlinearity which can be employed for all-optical variation of scattering behavior~\cite{Noskov}. However, plasmonic nanoantennas suffer from high Joule losses and heating, which limit the tuning capabilities of such systems.
Silicon, on the other hand, has become a promising platform for implementation of nonlinear photonic devices thanks to a broad range of optical nonlinearities such as Kerr effect, two-photon absorption, and electron-hole plasma (EHP) excitation~\cite{Leuthold}. Silicon nanoantennas demonstrate a damage threshold far exceeding that of their plasmonic counterparts, thus enabling higher degree of tuning.
Recently, enhancement of optical nonlinearities in silicon has been demonstrated at the scale of single nanoparticles \cite{Shcherbakov2014,Lewi2015, Makarov2015,Shcherbakov2015, yang2015nonlinear}. In particular, photoexcitation of EHP was employed for tuning of silicon nanoantenna optical properties in the IR and visible regions~\cite{Lewi2015,Makarov2015}. 

\begin{figure}[!b]
\includegraphics[width=1\columnwidth]{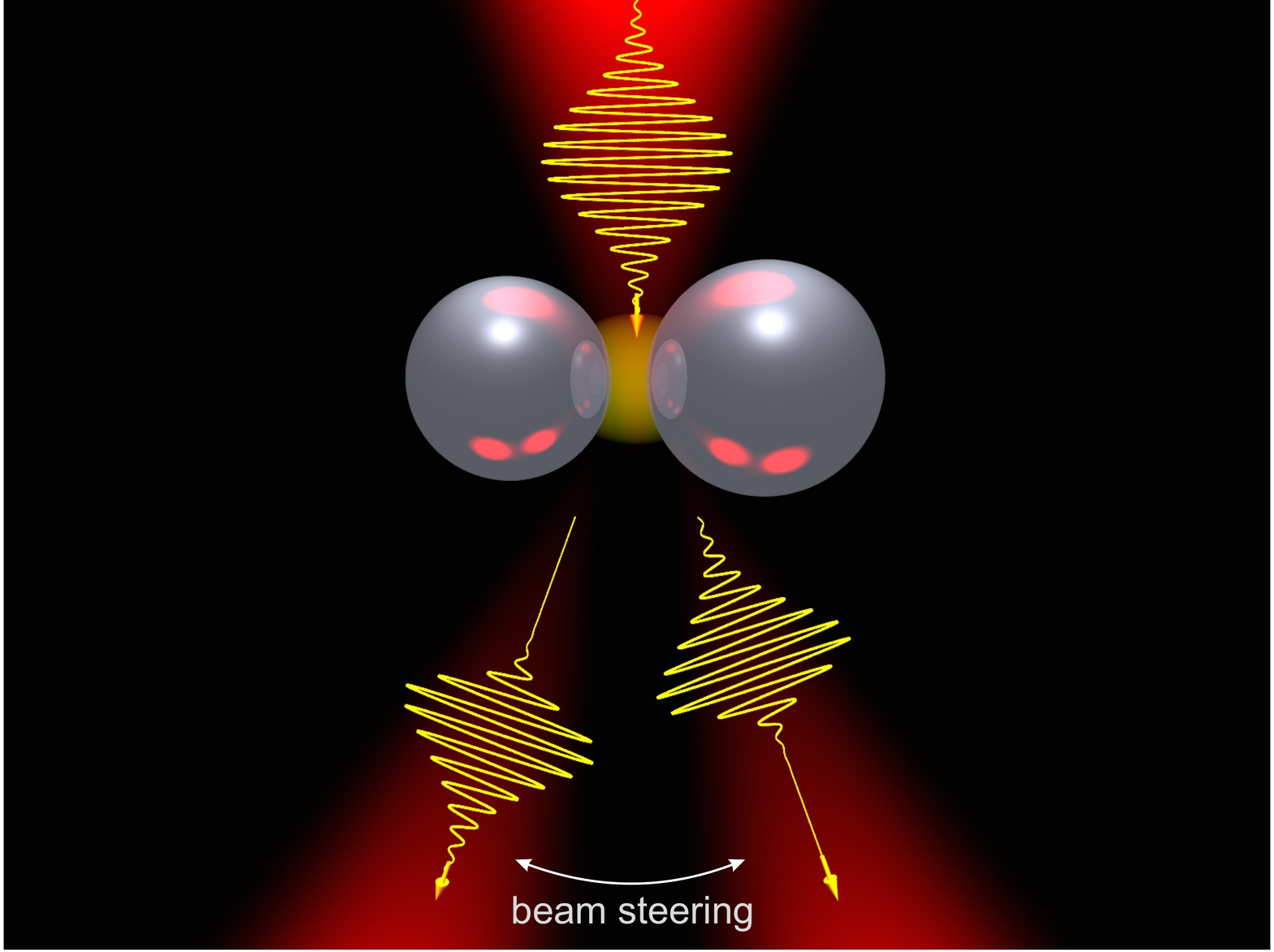}
\caption{A schematic view of beam steering via EHP photoexcitation in silicon nanoparticles.}
\label{fig1}
\end{figure} 

In this paper, we explore the capabilities of silicon nanoparticle dimers for nonlinear optical tuning enabled by photoexcitation of EHP.
In particular, we demonstrate nonlinear beam steering in an asymmetric dimer. The main direction of scattered light is controlled via intensity of incident pulse. For a realistic 200~fs pulse with a peak intensity of about 40~GW/cm$^2$ we observe steering of the scattering direction as large as 20 degrees compared to a weak reference pulse. Apart from the far-field properties of a nanoantenna manifested in its scattering diagram, we investigate how the near-field behavior, namely, local density of optical states (LDOS) can be controlled in the vicinity of the nanodimer via EHP excitation. We observe almost two-fold variation of LDOS in the dimer gap when 40~GW/cm$^2$ is applied. 
Our findings provide an additional tool for controlling light scattering at the nanoscale and prove the potential of silicon and germanium for the development of nanoscale all-optical devices.

\section{Model}
The proposed system for nonlinear beam steering as well as its operation principle are schematically shown in Fig.~\ref{fig1}. The two high-index dielectric nanoparticles of radii $R_1$ and $R_2$ surrounded by vacuum and separated by distance $L$ comprise the asymmetric dimer. Incident optical pulse enables photoexcitation of EHP within nanoparticles affecting their optical resonances and therefore scattering behavior. Silicon (Si) is chosen as a high-index material due to its high two-photon absorption at optical frequencies resulting in efficient EHP excitation~\cite{Plasma}. On the other hand, germanium (Ge) may become the optimal choice in the near-IR due to its attractive nonlinear properties.

\begin{figure}[!t]
\includegraphics[width=1\columnwidth]{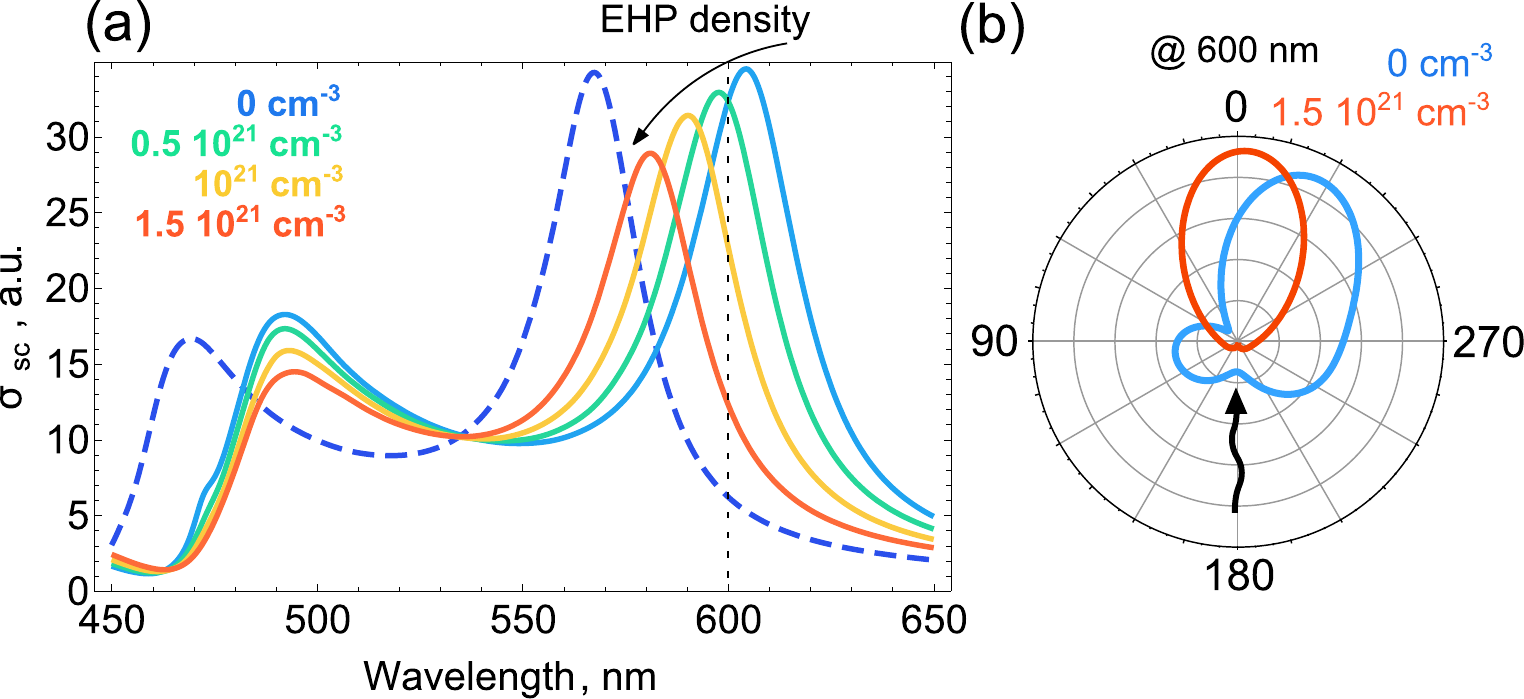}
\caption{ (a) Scattering cross-section of isolated nanoparticles with increasing EHP density in the resonant particle. Dashed curve shows cross-section of the non-resonant particle. (b) Scattering diagram for the asymmetric dimer for different EHP densities in the resonant nanoparticle. }
\label{fig2}
\end{figure}

In order to simulate nonlinear scattering of an optical pulse we adopt our analytical model developed in Ref.~\cite{Plasma}. Each spherical nanoparticle is modeled as a combination of electric ($\mathbf{p}$) and magnetic ($\mathbf{m}$) dipole moments. The temporal dynamics of slowly varying amplitudes of these dipoles is governed by oscillator equations:
\begin{equation}
\begin{gathered}
  i\frac{{\partial \alpha _e^{ - 1}}}{{\partial \omega }}\frac{{d{\mathbf{p}}}}{{dt}} + \alpha _e^{ - 1}{\mathbf{p}} = {{\mathbf{E}}_{{\text{inc}}}}\left( t \right), \hfill \\
  i\frac{{\partial \alpha _m^{ - 1}}}{{\partial \omega }}\frac{{d{\mathbf{m}}}}{{dt}} + \alpha _m^{ - 1}{\mathbf{m}} = {{\mathbf{H}}_{{\text{inc}}}}\left( t \right) \hfill \\ 
\end{gathered} 
\label{eq1}
\end{equation}
where dipole polarizabilities are expressed in terms of the Mie coefficients $a_1$ and $b_1$ as ${\alpha _e} = \frac{{3i}}{{2{k^3}}}{a_1}$ and ${\alpha _m} = \frac{{3i}}{{2{k^3}}}{b_1}$, respectively,\cite{Bohren} with $k=\omega/c$ being the free space wave vector. ${{\mathbf{E}}_{{\text{inc}}}}$ and ${{\mathbf{H}}_{{\text{inc}}}}$ denote slowly varying amplitudes of incident electric and magnetic fields.

The equations~(\ref{eq1}) fully describe dipole moments dynamics of a single nanoparticle provided that EHP density, which determines permittivity of photoexcited silicon, is known at all times. However, EHP excitation is driven by optical absorption within silicon due to the electric fields of induced dipoles. The dynamics of volume-averaged EHP density is described via the following rate equation~\cite{Plasma}:
\begin{equation}
\frac{{d{\rho _{{\text{eh}}}}}}{{dt}} =  - \Gamma {\rho _{{\text{eh}}}} + \frac{{{W_1}}}
{{\hbar \omega }} + \frac{{{W_2}}}{{2\hbar \omega }}.
\label{eq2}
\end{equation}
Here, $W_{1,2}$ are the volume-averaged dissipation rates due to one- and two-photon absorption, and $\Gamma$ is the phenomenological EHP relaxation rate constant which depends on EHP density~\cite{Cardona}. The absorption rates are written in the usual form as ${W_1} = \frac{\omega }{{8\pi }}  \left\langle {{{\left| {{\mathbf{\tilde E_{\rm in}}}} \right|}^2}} \right\rangle {\rm Im} (\varepsilon)$
 and ${W_2} = \frac{\omega }{{8\pi }} \left\langle {{{\left| {{\mathbf{\tilde E_{\rm in}}}} \right|}^4}} \right\rangle \operatorname{Im} {\chi ^{(3)}}$, where $\left\langle {...} \right\rangle $ denotes averaging over the nanoparticle volume, and $\operatorname{Im} {\chi ^{(3)}} = \frac{{\varepsilon {c^2}}} {{8\pi \omega }}\beta $ with $\beta$ being two-photon absorption coefficient. These averaged fields should be related to the instantaneous values of electric and magnetic dipole moments. This is done by integrating the total field of the two spherical harmonics corresponding to the given values of $\mathbf{ p}$ and $\mathbf{m}$.
The relaxation rate of EHP in c-Si is dominated by Auger recombination~\cite{Kerr}: $\Gamma  = \Gamma _{\text{A}} \rho_{\rm eh}^2$ with $\Gamma_{\rm A}=4 \cdot 10^{ - 31}$~s$^{-1}$cm$^6$ (Ref.~\cite{Shank}).
In germanium, EHP relaxation is again mediated by Auger mechanism with $\Gamma_{\rm A}=7 \cdot 10^{ - 33}$~s$^{-1}$cm$^6$ (Ref.~\cite{GeAuger}).

The system of equations (\ref{eq1}) and (\ref{eq2}) should be completed by the expression relating permittivity of excited material $\varepsilon$ to EHP density $\rho_{\rm eh}$. For silicon, this dependence is represented as the following expression:\cite{Sokolowski-Tinten2000, Makarov2015}
\begin{eqnarray}
{\varepsilon }\left( {\omega ,{\rho _{{\text{eh}}}}} \right) = \varepsilon_{\text{0}}+ \Delta\varepsilon_{\rm bgr}+ {\Delta\varepsilon _{{\text{bf}}}} + {\Delta\varepsilon _{{\text{D}}}}
\label{eq3}
\end{eqnarray}
where ${\varepsilon _{{\text{0}}}}$ is the permittivity of non-excited material, whereas $\Delta\varepsilon_{\text{bgr}}$, ${\Delta\varepsilon _{{\text{bf}}}}$, and $\Delta\varepsilon _{{\text{D}}}$ are the contributions from band gap renormalization, band filling, and Drude term. The detailed expressions for all contributions in Eq.~(\ref{eq3}) are given in \emph{Supporting Information}. 
Turning to permittivity of photoexcited germanium, we note that in the IR range it is dominated by Drude contribution, expression for which is adopted from Ref.~\cite{GeDrude} (see \emph{Supporting Information} for details).

\begin{figure*}[!t]
\includegraphics[width=2\columnwidth]{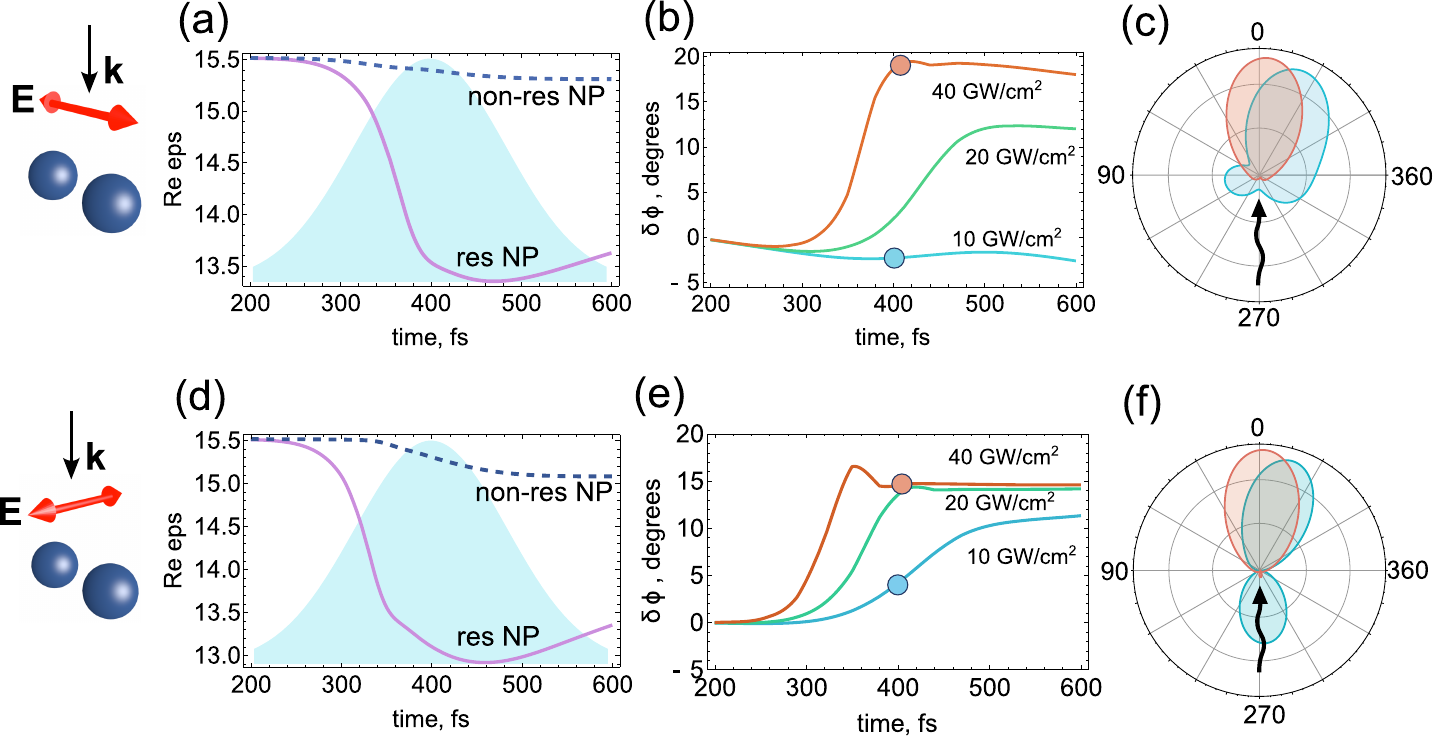}
\caption{(a) Time-dependent real part of permittivity of the two dimer nanoparticles under 40 GW/cm$^2$ pulse irradiation. Shaded area represents the incident pulse intensity. (b) The steering angle $\delta \phi$ as a function of time upon irradiation by pulses of different peak intensities. (c) The scattering diagrams of the photoexcited dimer corresponding to circles in panel (b). (d) - (f) The same as (a) - (c) but for $s$-polarized incident wave.}
\label{fig3}
\end{figure*}

When the nanoparticles form a nanodimer, the incident electric and magnetic fields ${{\mathbf{E}}_{{\text{inc}}}}\left( t \right)$, ${{\mathbf{H}}_{{\text{inc}}}}\left( t \right)$ in Eq.~(\ref{eq1}) must include the field of the incident plane wave as well as the local fields due to induced dipoles of the adjacent particle:
\begin{equation}
\begin{gathered}
  {\mathbf{E}}_{i.j}^{{\text{inc}}}\left( t \right) = {{\mathbf{E}}_0}\left( {{{\mathbf{r}}_{i,j}}} \right) + {k^2}{\mathbf{\hat G}}\left( {{{\mathbf{r}}_{i,j}},{{\mathbf{r}}_{j,i}}} \right){{\mathbf{p}}_{j,i}}\left( t \right) +  \hfill \\
  ik\nabla  \times {\mathbf{\hat G}}\left( {{{\mathbf{r}}_{i,j}},{{\mathbf{r}}_{j,i}}} \right){{\mathbf{m}}_{j,i}}\left( t \right), \hfill \\
  {\mathbf{H}}_{i.j}^{{\text{inc}}}\left( t \right) = {{\mathbf{H}}_0}\left( {{{\mathbf{r}}_{i,j}}} \right) + {k^2}{\mathbf{\hat G}}\left( {{{\mathbf{r}}_{i,j}},{{\mathbf{r}}_{j,i}}} \right){{\mathbf{m}}_{j,i}}\left( t \right) -  \hfill \\
  ik\nabla  \times {\mathbf{\hat G}}\left( {{{\mathbf{r}}_{i,j}},{{\mathbf{r}}_{j,i}}} \right){{\mathbf{p}}_{j,i}}\left( t \right), \hfill \\ 
\end{gathered} 
\label{eq5}
\end{equation}
Here ${\mathbf{\hat G}}\left( {{\mathbf{r}},{\mathbf{r'}}} \right) = \left( {{\mathbf{1}} + \frac{1} {{{k^2}}}\nabla  \otimes \nabla } \right)\frac{{{e^{ik\left| {{\mathbf{r}} - {\mathbf{r'}}} \right|}}}}
{{\left| {{\mathbf{r}} - {\mathbf{r'}}} \right|}}$ is the electric Green tensor, 
${{\mathbf{E}}_0}\left( {\mathbf{r}} \right) = \mathbf{E}_0{e^{i{\mathbf{kr}}}}$, 
${{\mathbf{H}}_0}({\mathbf{r}}) = {\mathbf{k}} \times {{\mathbf{E}}_0}({\mathbf{r}})/k$, and 
$\mathbf{k}$ is the incident wavevector.

The scattering pattern is determined via calculation of the time-averaged Poynting vector of the scattered field in the far-field zone:
\begin{equation}
{{\mathbf{S}}_{{\text{scat}}}} = \frac{c}
{{8\pi }}\operatorname{Re} \left( {{{\mathbf{E}}_{{\text{scat}}}} \times {\mathbf{H}}_{{\text{scat}}}^*} \right)
\end{equation}
where the scattered fields in the far-field zone are given by
\begin{equation}
\begin{gathered}
  {{\mathbf{H}}_{{\text{scat}}}} = {k^2}\sum\limits_j {{\mathbf{\hat G}}\left( {{\mathbf{r}},{{\mathbf{r}}_j}} \right){{\mathbf{m}}_j}}  - ik\sum\limits_j {\nabla  \times {\mathbf{\hat G}}\left( {{\mathbf{r}},{{\mathbf{r}}_j}} \right){{\mathbf{p}}_j}} , \hfill \\
  {{\mathbf{E}}_{{\text{scat}}}} \approx  - \frac{1}
{k}{{\mathbf{k}}_{{\text{scat}}}} \times {{\mathbf{H}}_{{\text{scat}}}}, \hfill \\ 
\end{gathered} 
\end{equation}
with $\mathbf{k}_{\rm scat}$ being wave vector in the scattering direction of interest.

\section{Beam steering}
We now apply our model to demonstrate the effect of efficient nonlinear beam steering. 
We consider scattering of an $p-$polarized optical pulse of $\lambda=600$~nm wavelength with its wave vector $\mathbf{k}$ normal to the dimer axis.
The dimer is formed by a magnetic dipole (MD) resonant particle of radius $R_1=74$~nm and non-resonant particle of radius $R_2=68$~nm separated by $L=220$~nm.
This geometry enables favorable conditions for beam steering (see \emph{Supporting Information}). Wavelength of 600 nm is chosen due to large two-photon absorption allowing to reduce intensities required for considerable switching.

The mechanism of steering can be understood by considering the optical properties of isolated particles comprising the dimer. The scattering cross-section spectra of the isolated particles in the linear regime are shown in Fig.~\ref{fig2}(a). The larger particle is tuned to the MD resonance at 600~nm, while the smaller particle is off resonance and close to the Kerker condition of unidirectional scattering, where $\alpha_{e}=\alpha_{m}$~\cite{Kerker}. This results in a strongly asymmetric radiation pattern of the dimer shown in Fig.~\ref{fig2}(b). The recent observation of directional scattering form silicon dimers~\cite{Maier15} suggests that the concept of EHP excitation in silicon nanoparticles can be employed for tailoring the radiation pattern at a constant wavelength but instead with varying incident intensity.

Indeed, when a strong pulse is incident on the dimer, it will enable dense EHP photoexcitation in the larger resonant particle, while the smaller non-resonant particle will be almost unaffected by the pulse. This will cause the MD resonance of the larger particle shift to shorter wavelengths due to refractive index decrease induced by EHP. Eventually, at certain level of photoexcited EHP density the effective resonance curves of the two particles may overlap leading to nearly symmetric forward scattering at $\lambda=600$~nm. This evolution of scattering patterns is shown in Fig.~\ref{fig2}(b) for different values of EHP density in the resonant particle. Such modification of optical response for single nanoparticles was recently demonstrated in Refs.~\cite{Makarov2015,Shcherbakov2015}.

This behavior is confirmed in numerical modeling of the transient nonlinear dynamics of the dimer. Time-dependent permirttivities of photoexcited Si in each of the two particles are shown in Fig.~\ref{fig3}(a) for a 200 fs Gaussian pulse with 40~GW/cm$^2$ peak intensity. 
Near the pulse center at $t \approx 400$~fs EHP induced permittivity correction in the resonant particle is nearly 5 times larger than that in the non-resonant particle. 
The attainable degree of steering depends on intensity of incident pulse. Moreover, since EHP excitation and relaxation are not instantaneous processes, it is useful to investigate dynamics of steering during the pulse action.
This dynamics for a series of peak intensities is presented in Fig.~\ref{fig3}(b), where the angle of the \emph{main lobe direction} $\delta \phi$ is shown as a funtion of time (the scattering direction in the cold regime is taken as 0).
Steering of the scattered radiation is maximal near the pulse center and slowly decreases afterwards owing to ps-scale EHP decay in silicon particles.
The corresponding scattering diagrams are shown in Fig.~\ref{fig3}(c), demonstrating nearly $20 ^\circ $ steering in comparison with the cold dimer.


Overall, similar dependencies are observed for $s$-polarized incident wave. 
The corresponding results for permittivity dynamics and scattering patterns are presented in Fig.~\ref{fig3}(d,e,f) assuming the same dimer geometry and incident pulse parameters as before. 
Again, only permittivity of the MD resonant particle is significantly affected, Fig.~\ref{fig3}(d).
A smaller value of steering of about $15 ^\circ $ degrees is observed for 40 GW/cm$^2$ pulse, Fig.~\ref{fig3}(e).
At the same time pulses of lower intensities result in larger steering in contrast to $p-$polarized incidence: here, 10 GW/cm$^2$ pulse produces $\approx 10^\circ $ degrees rotation of the main lobe, while for s polarization the scattering pattern is almost unaffected.
Unfortunately, there is significant backward scattering in the unexcited regime due to different picture of magnetic and electric dipoles interference.

\begin{figure}[!t]
\includegraphics[width=.9\columnwidth]{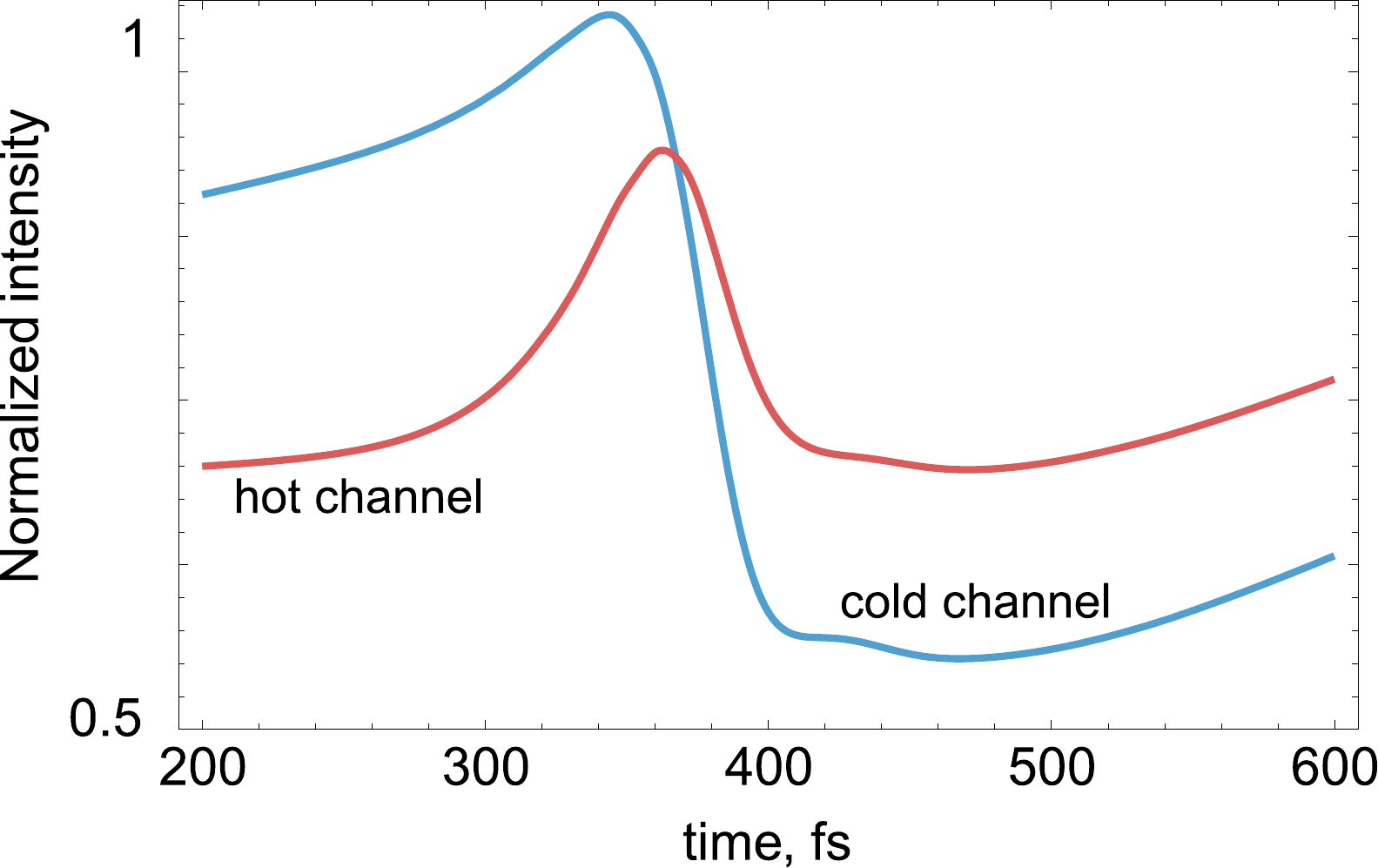}
\caption{Normalized intensity of radiation scattered by a Si nanodimer along the two directions as a function of time.}
\label{channel}
\end{figure}

In order to gain better impression of a nonlinear Si nanodimer as a light router, in Fig.~\ref{channel} we present time dependent scattered intensity (normalized by the instantaneous incident intensity) emitted in the two scattering channels: along the cold direction ($\delta \phi=0 ^\circ $) and along the forward direction ($\delta \phi=20 ^\circ $) for $p-$polarized 40~GW/cm$^2$ pulse. It clearly shows that plasma induced nonlinearity enables redistribution of the emitted radiation between the two channels with the relative difference of about 20\%.

\begin{figure}[!t]
\includegraphics[width=1\columnwidth]{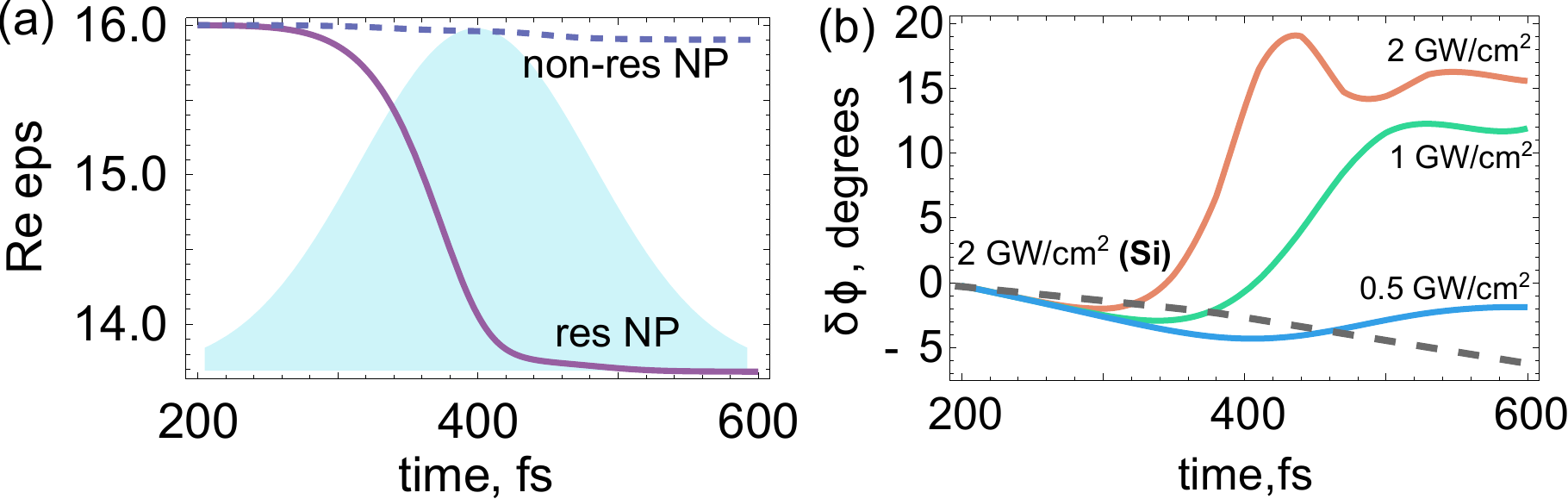}
\caption{(a) Time-dependent permittivity of photoexcited Ge in the two dimer particles irradiated by a 200 fs pulse at 2 $\mu$m wavelength. (b) The steering angle form a Ge dimer as a function of time for a 200 fs pulse with different peak intensities. Dashed curve shows steering from a Si dimer for the same incident pulse.}
\label{figGe}
\end{figure}


While silicon exhibits attractive characteristics for its use at optical frequencies, it has weak two-photon absorption in near-IR ($\beta<2$~cm/GW) and therefore is not preferable for tunable nanoantennas mediated by EHP excitation.
To extend the idea of EHP-controlled dimer nanoantenna beyond the visible region, we note that germanium is a promising candidate for all-dielectric nanophotonics in the near-IR~\cite{Ge2011}, where it is almost lossless with refractive index $\approx 4$. Remarkably, it shows huge two-photon absorption ($\beta \approx 80$~cm/GW at 2.9~$\mu$m, Ref. ~\cite{GeTPA}). Along with the fact that lower EHP densities are required for nanoantenna tuning at longer wavelengths due to increasing free electrons contribution to $\varepsilon$ (which scales as $1/\omega$), high TPA coefficient allows for decreasing of incident intensity required for beam steering.

The results for nonlinear beam steering in a Ge dimer at $\lambda=2~\mu$m are shown in Fig.~\ref{figGe} for $p-$polarized incident wave. Following the same strategy as before, we compose the dimer of a MD resonant particle (240 nm radius) and a particle obeying the Kerker condition (220 nm radius); the particles are separated by $L=800$~nm. In agreement with our expectations, 2 GW/cm$^2$ pulse of 200 fs duration causes excitation of EHP density in the resonant particle resulting in nearly $20 ^\circ$ degrees steering, Fig.~\ref{figGe}(b). To justify the choice of Ge over Si in near-IR, we show the steering angle in a Si dimer composed of MD resonant and Kerker particles (at 2 $\mu$m) in Fig.~\ref{figGe}(b), which demonstrates negligible tuning as compared to the Ge dimer.

\section{Near-field tuning}
Above we have demonstrated the effect of optical EHP excitation on the far field properties of the dimer nanoantenna, i.e., its scattering pattern. The plasma nonlinearity, however, can be also employed for tuning of the near field behavior, which can be characterized by the local density of states (LDOS) in the gap of the dimer.

\begin{figure}[!t]
\includegraphics[width=.9\columnwidth]{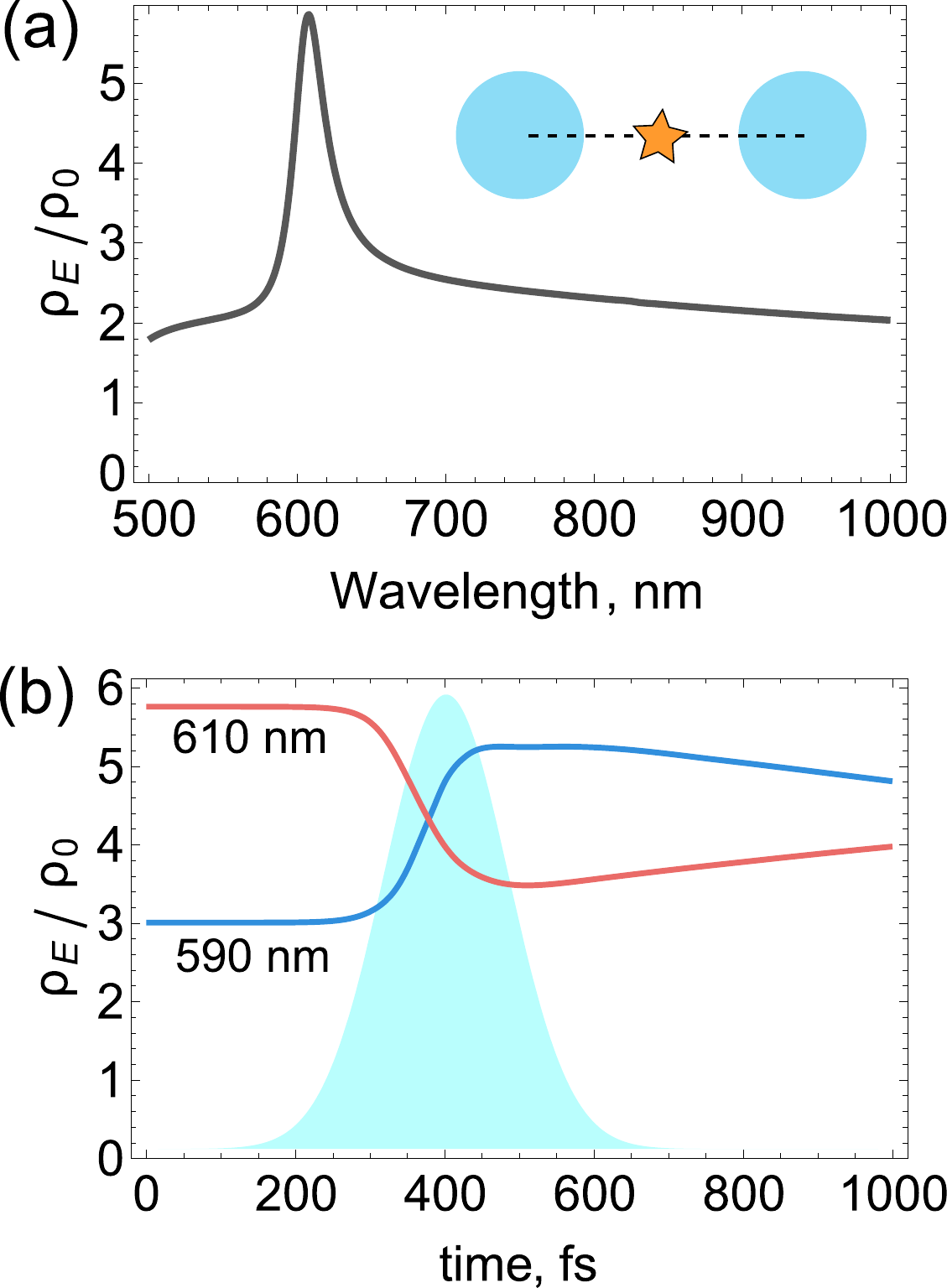}
\caption{(a) Spectrum of the orientation averaged electric LDOS ${\rho _{\text{E}}}$ at the dimer center in the unexcited state. (b) Electric LDOS as a function of time upon irradiation with a 40 GW/cm$^2$ pulse. Shaded area represents the incident pulse intensity.}
\label{fig5}
\end{figure}

Using the coupled dipoles approximation we calculate the orientation averaged electric LDOS 
${\rho _{\text{E}}} = \frac{{2\omega }}
{{3\pi {c^2}}} \times \operatorname{Im} \left( {{\text{Tr}}\left[ {{\mathbf{\hat G}}({{\mathbf{r}}_0},{{\mathbf{r}}_0};\omega )} \right]{\text{ }}} \right)$
 at the center of the symmetric Si dimer. 
The LDOS spectrum for a cold dimer without EHP is shown in Fig.~\ref{fig5}(a) exhibiting the well-known series of peaks associated with resonances of the dimer~\cite{Bakker2015,Albella2013}.
The time-dependent LDOS enhancement with respect to the free space LDOS ${\rho _0} = \frac{{{\omega ^2}}}{{3{\pi ^2}{c^3}}}$ is plotted in Fig.~\ref{fig5}(b) for different wavelengths assuming 200 fs pulse with 40 GW/cm$^2$ peak intensity as that used in the previous section for a Si dimer.

The results demonstrate that a 200 fs optical pulse can induce nearly 50\% change of LDOS at the dimer center for specific wavelength. 
After the pulse action LDOS returns to its initial value during the ps-scale EHP relaxation.
Our calculations also indicate that for a fixed transverse orientation of a dipole in the dimer gap the \emph{projected} LDOS $\rho_{E,\mathbf{u}}$ modification can be even more dramatic reaching five-fold enhancement or suppression.
 Importantly, photoexcitation of EHP can transform the system from enhanced state with $\rho_{E,\mathbf{u}}>\rho_{0}$ to the state with suppressed projected LDOS $\rho_{E,\mathbf{u}}<\rho_{0}$. 
Demonstrated tunability of LDOS in a silicon nanodimer allows for selective enhancement or suppression of various optical effects whose strength is dependent on LDOS. 
Those include not only the well-known Purcell effect, but also thermal emission~\cite{CarminatiReview} and nonlinear optical effects~\cite{Soljacic}.

\section{Discussion and Conclusion}
We finally add a remark regarding the modulation rate which may be attained with the proposed structure. Figures~\ref{fig3} and~\ref{fig5} demonstrate that the direct switching from the unexcited to the photoexcited state of a dimer takes $\sim 200$~fs, whereas Auger recombination in c-Si enables ultrafast recombination time down to sub-10-ps level for intensities higher than 20~GW/cm$^{2}$.
The fastest relaxation can be achieved with nc-Si nanoparticles, for which 2.5~ps EHP relaxation has been demonstrated~\cite{Plasma} resulting in approximately 400~Gbit/s bandwidth.
This large modulation speed may enable additional applications in the area of non-reciprocal emission and absorption at the nanoscale~\cite{Alu2, Alu3}.

To conclude, we have explored the potential of all-dielectric nanoparticle dimers for nonlinear manipulation of the near and far electromagnetic fields via electron-hole plasma photoexcitation.
In particular, we have demonstrated nonlinear steering of light scattered from an asymmetric dimer of silicon nanoparticles, where the steering angle is controlled via the intensity of incident optical pulse. 
Excitation with a 200 fs pulse of 40~GW/cm$^2$ peak intensity allows to achieve $20 ^\circ$ steering. 
The concept was also applied in the near-IR for a germanium dimer.
Plasma excitation also enables control of the near fields manifested in the local density of states in the vicinity of a dimer. 
We have shown that excitation of plasma can induces 50\% variation of orientation averaged LDOS and even more dramatic change of projected LDOS for a fixed orientation.
This variation may be employed for transient selective control of LDOS-sensitive effects such as spontaneous and thermal emission.

\begin{acknowledgments}
This work was supported by the Ministry of Education and Science of the Russian Federation (project No 14.584.21.0009 with unique identificator RFMEFI58414X0009). D.G.B. acknowledges support from the Russian Foundation for Basic Research (project No 16-32-00444).
\end{acknowledgments}

\bibliography{steering}

\end{document}